\documentclass[a4paper,12pt]{article}
\usepackage{amssymb,amsthm,amsmath,amsfonts}
\usepackage{float}
\usepackage{cite}
\usepackage[pdftex]{graphicx} 
\usepackage[spanish, english]{babel} 
\usepackage[utf8]{inputenc}
\usepackage[hyphens]{url}
\usepackage[breaklinks,colorlinks=true,linkcolor=black, citecolor=black, urlcolor=blue]{hyperref}
\usepackage{graphics}

\numberwithin{equation}{section}

\begin{document}

\title{{Condición de Lorentz y ecuaciones de ondas electromagnéticas como propiedades emergentes del sistema de Maxwell}}

\author{Yudier Pe\~na P\'erez$^{a_1}$, Juan Bory Reyes$^{a_2}$.}
\date {\small{{$^a${Instituto Politécnico Nacional, Escuela Superior de Ingeniería Mecánica y Eléctrica-Unidad Zacatenco, Sección de Estudios de Posgrado e Investigación. Ciudad de México, México.\\
$^{a_1}${e-mail: ypenap88@gmail.com}, $^{a_2}${e-mail: juanboryreyes@yahoo.com}\\
}}}}

\maketitle
\selectlanguage{english}
\begin{abstract}
En el presente artículo se estudian las ecuaciones de ondas electromagnéticas y la condición de Lorentz, como propiedades emergentes del sistema de Maxwell en el contexto de la Teoría de Sistemas. Para este fin, se deducen las ecuaciones de ondas y la ecuación de Helmholtz. Haciendo uso del operador de Dirac desplazado y su estrecha relación con los principales operadores del cálculo vectorial, es posible establecer una conexión directa entre las soluciones del sistema de Maxwell tiempo-armónico y dos ecuaciones cuaterniónicas. Además, se expone la aplicación de la condición de Lorentz para transformar el sistema de Maxwell tiempo-armónico en una simple ecuación cuaterniónica en función de los potenciales escalar y vectorial.
 
This article deals with the study of electromagnetic waves equations and the Lorentz condition, as emergent properties of Maxwell's system in the context of systems theory. To do this, the wave equations and the Helmholtz equation are first deduced. Using the displaced Dirac operator, which is closely related to the main vector calculation operators, it is possible to establish a direct connection between the solutions of the Maxwell time-harmonic system and two quaternion equations. Also, the application of the Lorentz condition to transform the time-harmonic Maxwell system into a simple quaternion equation based on the scalar and vector potentials is exposed. 

\vspace{0.3cm}

\small{
\noindent
\textbf{Keywords.} Maxwell system, Emerging Property, Wave Equation, Lorentz Condition, Dirac operator.\\
\noindent
\textbf{PACS.} 01.70.+w; 02.30.-f; 03.50.De.}  
\end{abstract}

\selectlanguage{spanish}

\section{Introducción}
El concepto de emergencia tiene su origen en 1862, cuando en el libro ``\textit{A system of logic}", del filósofo inglés John Stuart Mill, aparece la idea de que la interacción y yuxtaposición de las partes que conforman un sistema, resulta insuficiente para entender y explicar las propiedades del sistema. El nombre del concepto se deriva de ``\textit{emergere}", que en latín significa ``salir de", y se atribuye a los filósofos británicos Samuel Alexander, Conwy Lloyd Morgan, entre otros \cite{Alexander, Morgan}. Fue a partir de 1920 que se desarrolló una variante del enfoque reduccionista de la evolución cultural, denominada la ``filosofía emergentista".

La Ciencia de la Complejidad presenta distintos orígenes en muchas disciplinas, lo que ha llevado a ``una metodología interdisciplinaria para explicar la emergencia de ciertos fenómenos macroscópicos a través de las interacciones no lineales de los elementos microscópicos" \cite{Mainzer}. Dependiendo del área, es posible encontrar varias definiciones de emergencia, pero en general, el término ``propiedad emergente" \hspace{0.02cm} hace alusión a cierta propiedad del sistema que difiere de la propiedad de cada componente individual y que resulta de las interacciones de tales componentes.
 
Según \cite{Aguilar}, no existe un consenso sobre el concepto de emergencia, éste no es compatible con el enfoque tradicional de la ciencia y hasta hoy no existe una justificación apropiada para el proceso de emergencia. No obstante, el concepto de emergencia reaparece frecuentemente a partir del siglo XIX. La emergencia implica un cambio de paradigma, y su aprobación por la comunidad científica resulta un importante desafío epistemológico.

No obstante, el autor de \cite{Aguilar} describe y analiza el concepto de emergencia con cierto rigor y profundidad, e investiga su relación con diferentes áreas de la ciencia. De igual modo, analiza conceptos vecinos fundamentales que ayudan a complementar la emergencia, y ofrecen algunas aplicaciones actuales que resultan muy útiles al resolver problemas de gran complejidad, además de permitir una mejor comprensión de Sistemas Complejos del mundo real.

En 1972, el físico Philip Anderson, publicó el trabajo ``\textit{More is Different}'' con lo que devolvió el emergentismo a la mesa teórica, abordando éste de una manera concisa, al observar que en cada nivel de complejidad aparecen propiedades nuevas \cite{Anderson}. 

El presente artículo se enfoca en el estudio de ciertas propiedades emergentes en el sistema de Maxwell, desde el punto de vista de la Teoría de Sistemas \cite{Goldstein,Sawyer,Bunge}, para ello se presentan la deducción de las ecuaciones de ondas electromagnéticas y la condición de Lorentz como propiedades emergentes del modelo del electromagnetismo dado originalmente por Maxwell. Además, haciendo uso de herramientas matemáticas, en particular del Análisis Cuaterniónico; y de las relaciones entre los campos electromagnéticos y los potenciales escalar y vectorial, se llegará a dos reformulaciones diferentes del sistema de ecuaciones de Maxwell tiempo-armónico. Lo anterior evidencia la aplicación e importancia de las propiedades que emergen en los Sistemas Complejos al ser considerados de manera global y no cada elemento de manera individual, como es el caso de las ecuaciones de ondas y la condición de Lorentz en el sistema de ecuaciones de Maxwell \cite{Kirsch}.  

La estructura del artículo es como sigue: la Sección 2 muestra algunos elementos básicos de Análisis Cuaterniónico que serán utilizados en el trabajo. La Sección 3 describe conceptos de emergencia en la Teoría de Sistemas. En la Sección 4 se ilustra la deducción y aplicación de las propiedades emergentes objeto de estudio. Finalmente, en la Sección 5 unas breves conclusiones.

\section{Bases del Análisis Cuaterniónico}
El Análisis Cuaterniónico fue iniciado por Rudolf Fueter en \cite{Fueter}. Luego, otros autores se interesaron en establecer nuevos enfoques para desarrollar las ideas básicas de esta teoría \cite{Haefeli, Deavours, Sudbery}.

Sea $\mathbb H(\mathbb R)$ el cuasicampo de los cuaternios reales y sean $e_0=1, e_1, e_2, e_3$ las unidades cuaterniónicas que cumplen con las reglas de multiplicación:
\[e_ie_j+e_je_i=-2\delta_{ij},\,\,i,j=1,2,3,\]
\[e_1e_2=e_3,\,\,e_2e_3=e_1,\,\,e_3e_1=e_2.\]

Para cada cuaternio $a=a_0+\vec{a}=\sum_{j=0}^{3}a_je_j$, $a_0:=Sc(a)$ se denomina parte escalar mientras que $\vec{a}:=Vec(a)$ es la parte vectorial del cuaternio $a$. Si $Sc(a)=0$, el cuaternio es denominado puramente vectorial y se identifica con el vector $\vec{a}=(a_1,a_2,a_3)$ del espacio $\mathbb R^3$.
La multiplicación de dos cuaternios arbitrarios $a,b$ puede ser escrita considerando sus partes escalares y vectoriales, como sigue:
\[
ab=a_0b_0-\vec{a}\cdot\vec{b}+a_0\vec{b}+b_0\vec{a}+\vec{a}\times\vec{b}.
\]
Aquí y en adelante, $\vec{a}\cdot\vec{b}$ denota el producto escalar mientras que $\vec{a}\times\vec{b}$ denota el producto cruz usual, ambos en $\mathbb R^3$.
El álgebra de los cuaternios complejos, la cual se denota por $\mathbb H(\mathbb C)$, resulta de considerar cuaternios cuyas componentes son todas números complejos.
Consideraremos funciones definidas en conjuntos abiertos de $\mathbb R^3$ y tomando valores en $\mathbb H(\mathbb C)$. Estas funciones pueden escribirse como $u=\sum_{j=0}^{3}u_je_j$, con $u_j$ tomando valores complejos. Si las funciones componentes de la función $u$ son continuas, diferenciables, etc., las funciones $u$ serán, por definición, también continuas, diferenciables, etc.

Sea $D$ el operador de Dirac (Moisil-Teodorescu):   
\[D=\sum_{j=1}^{3}e_{j}\partial_{ x_{j}}, \,\,\,\,\, \partial_{ x_{j}}=\frac{\partial}{\partial x_{j}}.\]
Es conocido que $D^2=-\Delta$, el negativo del Laplaciano en $\mathbb R^3$.

Una función $u$ que toma valores en ${\mathbb H(\mathbb C)}$, definida y diferenciable en un abierto $\Omega\subset\mathbb R^3$, se denomina hiperholomorfa en $\Omega$ si y sólo si $D u=0$ en $\Omega$.  

Si $u=\sum_{j=0}^{3}u_je_j=u_{0}+\vec{u}$, la acción del operador $D$ sobre $u$ se escribe en forma vectorial como
\begin{equation}\label{Dirac vectorial}
D u=-\textup{div}\vec{u}+\textup{grad}u_{0}+\textup{rot}\vec{u}.
\end{equation}

El operador de Helmholtz resulta de gran importancia e interés, ya que aparece en distintas aplicaciones en Física. Para $\kappa\in\mathbb C$, éste puede ser factorizado como
\begin{equation}\label{DiracFactorizado}
\Delta+\kappa^2=-(D+\kappa)(D-\kappa).  
\end{equation}

Una función $u$ que toma valores en ${\mathbb H(\mathbb C)}$ se dice que es $\kappa$-hiperholomorfa en $\Omega\subset\mathbb R^3$ si
\begin{equation*}
D_{\kappa}u:= Du+\kappa u=0\,\,\,\textup{en}\,\,\,\Omega.
\end{equation*}

\section{Propiedades emergentes}
\begin{figure}[H]
\centering
\includegraphics[width=8cm,height=6cm]{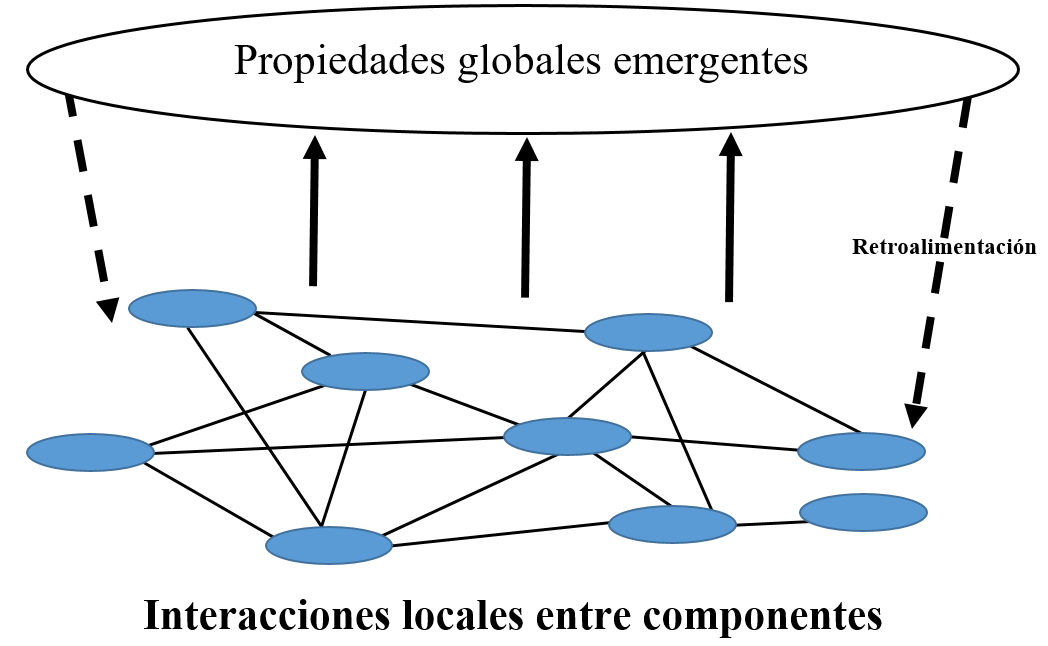}
\caption{Emergencia Global}
Fuente: \cite{Caparrini} 
\label{EmergenciaGlobal}
\end{figure}

Algunos investigadores de las Ciencias de la Complejidad \cite{Waldrop, Holland} llaman ``emergentes'' a fenómenos que surgen en los Sistemas Complejos. La emergencia y la complejidad se refieren al surgimiento de propiedades y comportamientos de nivel superior de un sistema que obviamente provienen de la dinámica colectiva de sus componentes \cite{Aziz1} (Figura \ref{EmergenciaGlobal}). 

En la actualidad, gran parte de los intentos para definir el concepto de emergencia, se apoyan en el pensamiento holístico (``el todo'' es siempre más que la suma de ``las partes'') o en el análisis del carácter que presentan las interrelaciones entre ``las partes'' y ``el todo''; así como el carácter de las propiedades que adquieren las ``partes'' y el ``todo''. Por ejemplo, una molécula de aire no es un ciclón y una especie aislada no forma una cadena alimentaria \cite{Aziz2}.

Bedau en \cite{Bedau}, define tres tipos de emergencia. Una emergencia nominal, cuando los ``fenómenos emergentes a nivel macro dependen de los fenómenos a nivel micro (los enteros dependen de sus constituyentes), y los fenómenos emergentes son autónomos de los fenómenos subyacentes (las propiedades emergentes no se aplican a las entidades subyacentes)''.

Por el contrario, en la emergencia fuerte las propiedades emergentes tienen un poder causal irreducible en las entidades subyacentes. Es decir, fenómenos emergentes (fuertes) logran adquirir capacidades causales novedosas, lo que permite que los sistemas ejerzan algún efecto de arriba hacia abajo.

Según la emergencia débil, características sistémicas en el nivel ``superior'' no son predecibles aun conociendo las leyes y características que gobiernan cada componente del sistema. 

En \cite{Fromm} se hace mención a las dos corrientes filosóficas que estudian los Sistemas Complejos. La filosofía tradicional del reduccionismo es ``encontremos primero las partes y leyes más fundamentales''. La filosofía complementaria, el emergentismo, es ``descubramos primero cómo surge la complejidad en los Sistemas Complejos, cómo surgen los fenómenos macroscópicos de las interacciones microscópicas''.

\begin{figure}[H]
\centering
\includegraphics[width=8.5cm,height=7cm]{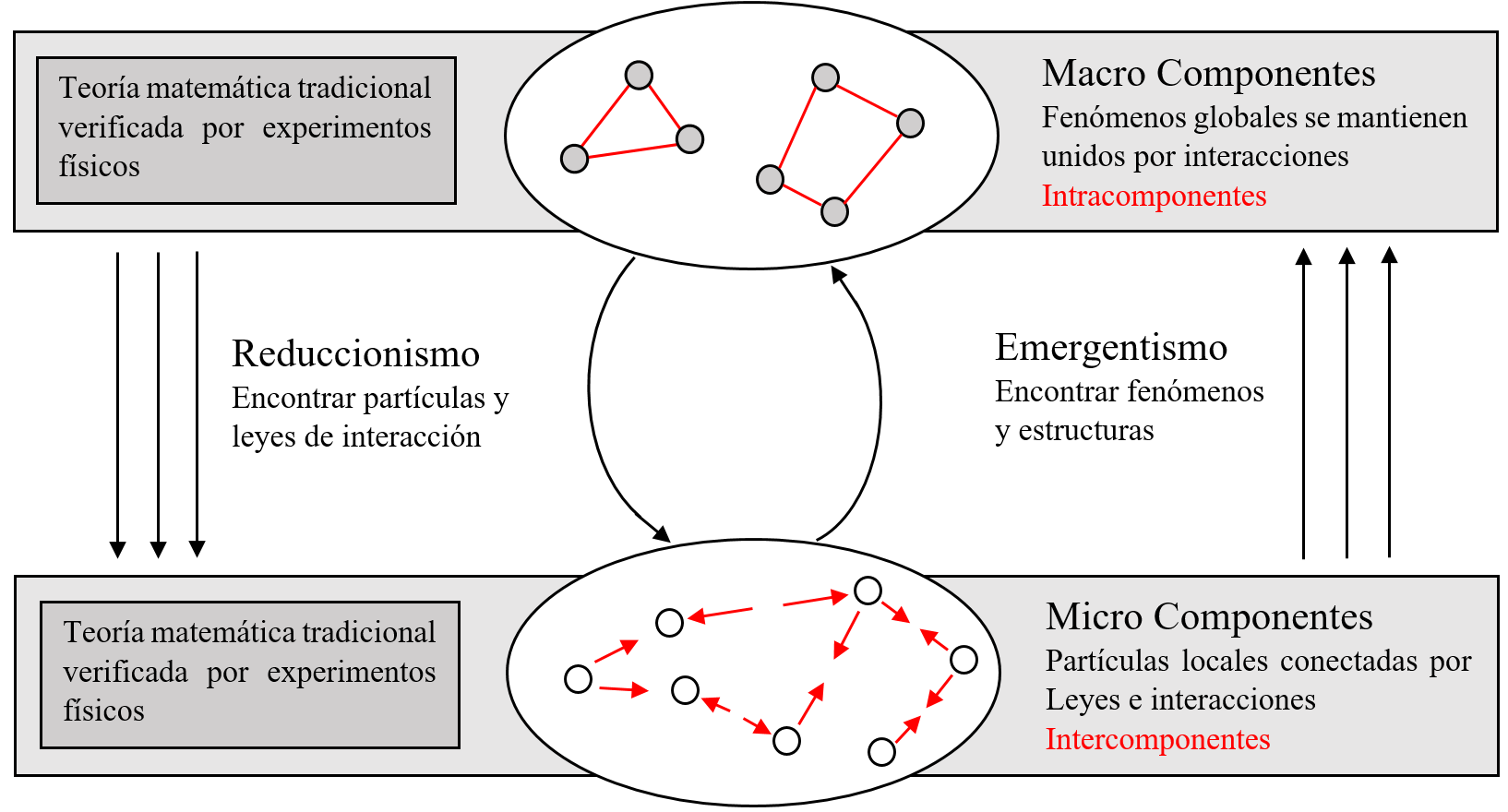}
\caption{Reduccionismo y Emergencia}
Fuente: \cite{Fromm} 
\label{Reduccionismo}
\end{figure}

Ambas direcciones son importantes, el reduccionismo y el emergentismo, éste último ayuda a comprender la unión entre el comportamiento micro y macroscópico. El descubrimiento del átomo y otras partículas fundamentales y las teorías correspondientes hicieron posible las Ciencias Naturales. Pero si no conocemos qué tipo de fenómenos macroscópicos pueden surgir de estos elementos microscópicos, partículas y leyes, entonces este conocimiento es desarticulado y no coherente (Figura \ref{Reduccionismo}).
Ambas corrientes son complementarias entre sí, el emergentismo necesita una base, y el reduccionismo necesita conexión y coherencia: el emergentismo sin reduccionismo es vago y poco claro, el reduccionismo sin el emergentismo no está conectado y no es coherente.

En la Teoría General de Sistemas \cite{Bertalanffy, Boulding}, propiedad emergente significa que, mediante arreglos específicos, un sistema exhibe propiedades o comportamientos que ninguno de sus componentes posee. Quizás el ejemplo más simple y llamativo es el agua. Tanto el hidrógeno como el oxígeno son gases en su origen natural, pero al mezclarlos convenientemente, se obtiene una sustancia líquida: el agua \cite{Aziz2}.

Goldstein en \cite{Goldstein}, antes de dar una definición, realiza una caracterización: ``la emergencia no funciona tanto como explicación, es más bien un término descriptivo que señala los patrones, estructuras o propiedades que surgen a nivel macro''. A su juicio, la emergencia se refiere al “surgimiento de estructuras, patrones y propiedades nuevas y coherentes durante el proceso de autoorganización en los sistemas complejos”.

Resulta particularmente consecuente con la perspectiva generativa la definición de Sawyer en \cite{Sawyer}: ``un sistema exhibe emergencia cuando a nivel macro se dan propiedades emergentes permanentes que surgen en forma dinámica de las interacciones entre las partes en el nivel micro''.

La definición de Bunge \cite{Bunge}, es consecuente con su tendencia a la formalización: ``otra categoría filosófica resaltada por el enfoque sistémico es la de emergencia; se dice que cierta propiedad de un sistema es emergente en el nivel N si ninguna de las partes de N la posee''.

En resumen, los enfoques anteriores definen las propiedades emergentes en un sistema, como \textit{``las características que no son posibles deducir al sumar cada una de las características de sus elementos de manera individual, sino que se hace necesario considerar las interrelaciones entre los elementos''}.

\section{Propiedades emergentes en el sistema de ecuaciones de Maxwell}
Lo que se entiende por emergencia es posible considerarlo de distintos modos \cite{Vivanco}, y su construcción desde el punto de vista teórico, resulta un novedoso campo para la obtención de respuestas y explicaciones \cite{Johnson}. A continuación se estudian propiedades emergentes del sistema de Maxwell en un contexto sistémico, de acuerdo a los enfoques antes mencionados. 

Los fenómenos electromagnéticos se describen con la ayuda de las funciones $\bf E$ (campo eléctrico) y $\bf B$ (inducción magnética) definidas en $\mathbb R_{\vec{x}}^3\times \mathbb R_t$ con valores vectoriales en $\mathbb R^3$.\\
$\bf E$ y $\bf B$ están vinculadas con las funciones ${\bf j}$ y $\rho$ también definidas en $\mathbb R_{\vec{x}}^3\times \mathbb R_t$ (con la densidad de corriente ${\bf j}(\vec{x},t)\in \mathbb R^3$ y densidad de carga $\rho(\vec{x},t)\in \mathbb R$) por las ecuaciones de Maxwell \cite{Kirsch}: 
\begin{equation}\label{Ampere}
\textup{rot}\,\, {\bf B}=\mu\varepsilon\displaystyle\frac{\partial {\bf E}}{\partial t}+\mu\bf j,\,\,\,\,\,\,\,\,\,\, \textup{(Ley de Ampere)},
\end{equation}
\begin{equation}\label{Faraday}
\textup{rot}\,\, {\bf E}=-\displaystyle\frac{\partial {\bf B}}{\partial t},\,\,\,\,\,\,\,\,\,\,\,\,\,\,\,\,\,\,\,\,\,\,\,\,\, \textup{(Ley de Faraday)},
\end{equation}
\begin{equation}\label{Electrica de Gauss}
\textup{div}\,\, \bf E=\frac{\rho}{\varepsilon},\,\,\,\,\,\,\,\,\,\,\,\,\,\,\,\,\,\, \textup{(Ley el\'ectrica de Gauss)},
\end{equation}
\begin{equation}\label{Magnetica de Gauss}
\textup{div}\,\, {\bf B}=0,\,\,\,\,\,\,\,\,\,\,\,\,\,\,\, \textup{(Ley magn\'etica de Gauss)},
\end{equation}
donde $\varepsilon$ y $\mu$ representan la permitividad y permeabilidad absolutas del medio, respectivamente.

\subsection{Ecuaciones de ondas}
En el caso de un medio simple no conductor con ausencia de fuentes, las ecuaciones (\ref{Ampere}-\ref{Magnetica de Gauss}) se escriben como:
\begin{equation}\label{Ampere1}
\textup{rot}\,\, {\bf B}=\mu\varepsilon\displaystyle\frac{\partial {\bf E}}{\partial t},
\end{equation}
\begin{equation}\label{Faraday1}
\textup{rot}\,\, {\bf E}=-\displaystyle\frac{\partial {\bf B}}{\partial t},
\end{equation}
\begin{equation}\label{Electrica de Gauss1}
\textup{div}\,\, {\bf E}=0,
\end{equation}
\begin{equation}\label{Magnetica de Gauss1}
\textup{div}\,\, {\bf B}=0.
\end{equation}
Tomando como base el sistema anterior, podemos obtener las ecuaciones de ondas \cite{Kirsch}, para ello el operador rotacional es aplicado en ambos lados de las ecuaciones (\ref{Ampere1}) y (\ref{Faraday1}), y se obtiene
\begin{equation}\label{EcuacionOndas1}
\begin{cases} 
\vspace{0.1cm}\textup{grad}(\textup{div}\,\,{\bf B})-\Delta {\bf B}=\mu\varepsilon\displaystyle\frac{\partial}{\partial t}[\textup{rot}\,\, {\bf E}],\cr\textup{grad}(\textup{div}\,\,{\bf E})-\Delta {\bf E}=-\displaystyle\frac{\partial}{\partial t}[\textup{rot}\,\, {\bf B}].
\end{cases}
\end{equation} 
De acuerdo con (\ref{Electrica de Gauss1}) y (\ref{Magnetica de Gauss1}) la divergencia de $\bf E$ y $\bf B$ es cero, y sustituyendo (\ref{Ampere1}) y (\ref{Faraday1}) en (\ref{EcuacionOndas1}), obtenemos las ecuaciones de ondas para los campos eléctrico y magnético.  
\begin{equation}\label{EcuacionOndas2}
\begin{cases} 
\vspace{0.3cm}\Delta {\bf E}-\displaystyle\frac{1}{c^2}\displaystyle\frac{\partial^2}{\partial t^2}[{\bf E}]=0,\cr\Delta {\bf B}-\displaystyle\frac{1}{c^2}\displaystyle\frac{\partial^2}{\partial t^2}[{\bf B}]=0.
\end{cases}
\end{equation}
Aquí y en adelante $c=\displaystyle\frac{1}{\sqrt{\mu\varepsilon}}$ denota la velocidad de propagación de las ondas.\\ 
Los campos $\bf E$ y $\bf B$ dependen de la posición y el tiempo, en el caso tiempo-armónico (teniendo una solución temporal $e^{i\omega t}$ con una frecuencia de oscilaciones $\omega$), de (\ref{EcuacionOndas2}) obtenemos las ecuaciones vectoriales de Helmholtz
\begin{equation}\label{EcuacionHelmholtz}
\begin{cases} 
\vspace{0.1cm}\Delta \vec{E}+\kappa^2 \vec{E}=0,\cr\Delta \vec{B}+\kappa^2 \vec{B}=0,
\end{cases}
\end{equation} 
donde $\kappa$ es el número de onda.

La ecuación de Helmholtz está estrechamente relacionada con el sistema de Maxwell (para campos armónicos en el tiempo). Las soluciones de la ecuación de Helmholtz se utilizan para generar soluciones del sistema de Maxwell (potenciales de Hertz), y cada componente del campo eléctrico y magnético satisface una ecuación de tipo Helmholtz. Por tanto, un procedimiento usual en el tratamiento del sistema de ecuaciones de Maxwell ha consistido en reducirlo a una ecuación equivalente de tipo Helmholtz \cite{Colton}.

\begin{figure}[H]
\centering
\includegraphics[width=8.5cm,height=6.5cm]{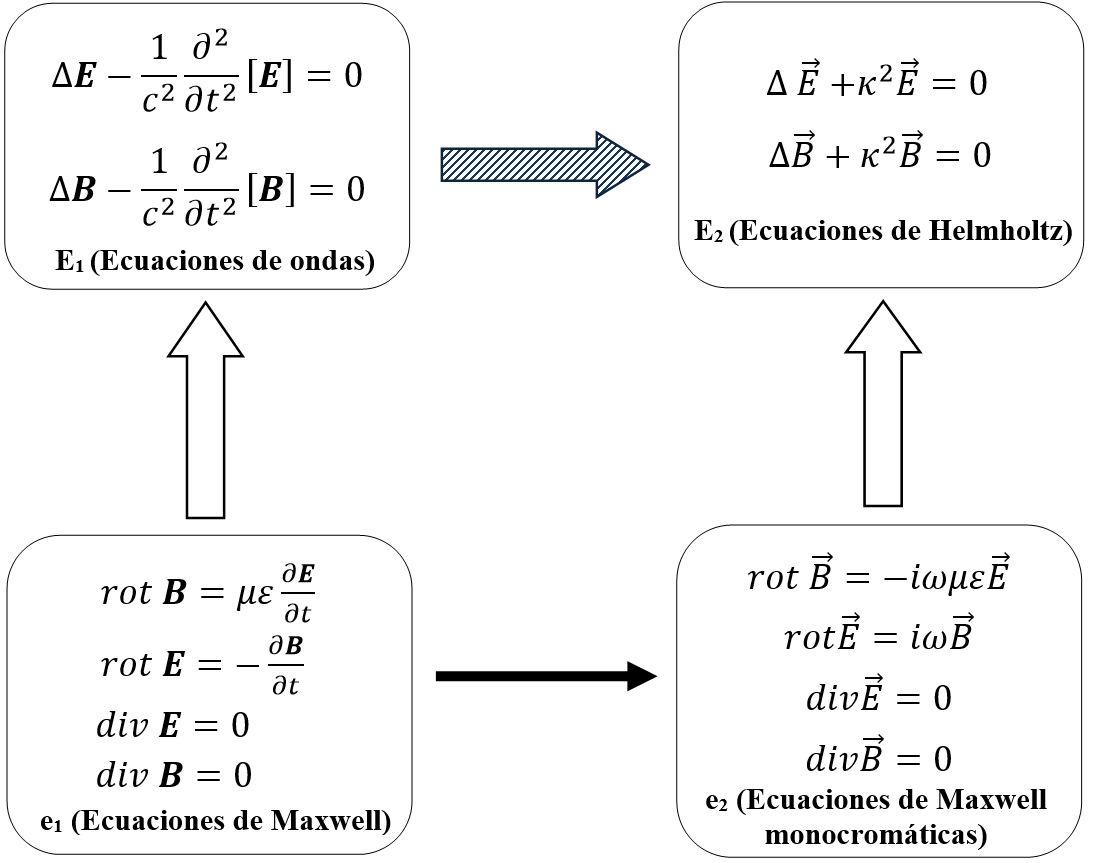}
\caption{Ecuaciones de ondas y niveles de descripción.}
Fuente: Elaboración propia 
\label{ModeloPropEmergentes}
\end{figure}

En la Figura \ref{ModeloPropEmergentes} se representan las ecuaciones de ondas como propiedad emergente de las ecuaciones de Maxwell, desde el punto de vista de la Teoría de Sistemas. Se considera el paso de las ecuaciones de Maxwell (estado inferior izquierdo) a las ecuaciones de Maxwell monocromáticas (estado inferior derecho) \cite{Fuentes}.
Inicialmente, en el nivel inferior el sistema se encuentra en el estado $e_1$ (ecuaciones de Maxwell), en tanto que para la presentación en el nivel superior se le designa el estado $E_1$ (ecuaciones de ondas). Como no es posible predecir $E_1$ a partir de considerar cada elemento de $e_1$ de manera individual, y para obtener el primero se hace necesario tomar en cuenta las interrelaciones de los elementos de $e_1$, entonces se dice que $E_1$ es un estado o propiedad emergente. De igual forma, si en el escenario que se encuentra el sistema en el estado $e_2$ (ecuaciones de Maxwell monocromáticas) se consideran las interrelaciones de sus elementos, entonces se obtiene el estado $E_2$ (ecuaciones de Helmholtz). De ahí que las flechas hacia arriba en la Figura \ref{ModeloPropEmergentes} indican emergencia \cite{Fuentes}.      

Las ecuaciones de Helmholtz (\ref{EcuacionHelmholtz}) motivaron la introducción del operador de Helmholtz $\Delta+\kappa^{2}I$ antes mencionado, y que puede ser factorizado como (\ref{DiracFactorizado}), utilizando el operador de Dirac desplazado $D_{\kappa}$.

Siguiendo las ideas de \cite{Kravchenko}, la relación mostrada en la ecuación (\ref{Dirac vectorial}) permite obtener una reformulación cuaterniónica del sistema de Maxwell tiempo armónico en $\mathbb R^3$. Dicha equivalencia provee una estructura algebraica más simple del sistema de Maxwell en la forma 
\begin{equation}\label{ReformulacionCuaternionica}
\begin{cases} 
D_{-\alpha_0}[\vec{\vartheta}]={\rm div}\vec{j}+\alpha_0\,\vec{j},\cr D_{\alpha_0}[\vec{\eta}]=-{\rm div}\vec{j}+\alpha_0\,\vec{j},
\end{cases}
\end{equation}
donde $\vec{\vartheta}=-i\omega\varepsilon\vec{E}+\alpha_0\vec{H}$, $\vec{\eta}=i\omega\varepsilon\vec{E}+\alpha_0\vec{H}$ son funciones puramente vectoriales que toman valores en ${\mathbb H} ({\mathbb C})$ (campos de Beltrami) y el número de onda $\alpha_0=\omega\sqrt{\varepsilon\mu}$ es seleccionado tal que $\mbox{Im}\, \alpha_0\geq0$. Es interesante hacer notar que $\sqrt{\varepsilon\mu}$ constituye el inverso de la velocidad de propagación de las ondas electromagnéticas en el medio.

\subsection{Condición de Lorentz y ecuaciones de ondas potenciales}
Con frecuencia, en el estudio de las ecuaciones de Maxwell microscópicas, las funciones $\bf E$ y $\bf B$ son sustituidas por las siguientes:  
\begin{equation}\label{Potencial vectorial}
(\vec{x},t)\rightarrow {\bf A}(\vec{x},t)\in \mathbb R^3,\,\,\,\,\,\,\,\,\,\,\,\,\,\,\,\,\, \textup{(Potencial vectorial)},
\end{equation}
\begin{equation}\label{Potencial escalar}
(\vec{x},t)\rightarrow V(\vec{x},t)\in \mathbb R,\,\,\,\,\,\,\,\,\,\,\,\,\,\,\,\,\,\,\,\,\,\,\,\, \textup{(Potencial escalar)},
\end{equation}
que están relacionadas con $\bf E$ y $\bf B$ por 
\begin{equation}\label{Potentialrelation}
\begin{cases} 
{\bf B}= \textup{rot}\,{\bf A},\,\,\,\,\,\,\,\,\,\cr{\bf E}=-\textup{grad}\,V-\displaystyle\frac{\partial {\bf A}}{\partial t}.
\end{cases}
\end{equation}
Sustituyendo en el sistema (\ref{Ampere}-\ref{Magnetica de Gauss}), obtenemos el siguiente sistema lineal no homogéneo:
\begin{equation}\label{SistemaLinealNohomogeneo}
\hspace{-0.1cm} 
\begin{cases}
\displaystyle\frac{1}{c^2}\frac{\partial^2 {\bf A}}{\partial t^2}-\Delta {\bf A} + \textup{grad}\left(\textup{div}\,{\bf A}+\displaystyle\frac{1}{c^2}\frac{\partial V}{\partial t}\right)= \mu{\bf j},\cr\,-\Delta V-\displaystyle\frac{\partial}{\partial t}\textup{div}\,{\bf A}=\displaystyle\frac{\rho}{\varepsilon}.
\end{cases}
\end{equation}
Observemos que las funciones ${\bf A}$ y $V$ no están definidas de manera única por (\ref{Potentialrelation}) a partir de $\bf E$ y $\bf B$: si ${\bf A}$ y $V$ satisfacen (\ref{Potentialrelation}), entonces para cualquier función arbitraria $u$ de $\vec{x}$ y $t$, ${\bf A^{'}}$ y $V^{'}$ definidos por:
\begin{equation}\label{PotentialrelationPrime}
\begin{cases} 
\vspace{0.2cm}{\bf A^{'}}= {\bf A}+\textup{grad}\,u,\,\,\,\,\,\,\,\,\,\cr V^{'}=V-\displaystyle\frac{\partial {u}}{\partial t},
\end{cases}
\end{equation}
también satisfacen (\ref{Potentialrelation}).\\
La transformación (\ref{PotentialrelationPrime}) es llamada transformación Gauge \cite{Arbab}, y al utilizarla obtenemos 
\begin{equation}\label{PotentialrelationPrime1}
\textup{div}\,{\bf A^{'}}+\frac{1}{c^2}\frac{\partial V^{'}}{\partial t}=\textup{div}\,{\bf A}+\frac{1}{c^2}\frac{\partial V}{\partial t}+\Delta u-\frac{1}{c^2}\frac{\partial^2 u}{\partial t^2}.
\end{equation}
Sea $u$ una solución de la ecuación
\begin{equation}\label{Solucionu}
\Delta u-\frac{1}{c^2}\frac{\partial^2 u}{\partial t^2}=-\left(\textup{div}\,{\bf A}+\frac{1}{c^2}\frac{\partial V}{\partial t}\right),
\end{equation}
(donde ${\bf A}$ y $V$ son conocidas), entonces es posible seleccionar un par $({\bf A_L},V_L)$ tal que
\begin{equation}\label{LorentzCondition}
\textup{div}\,{\bf A_L}+\frac{1}{c^2}\frac{\partial V_L}{\partial t}=0.
\end{equation}
Esta relación es llamada la condición de Lorentz \cite{Jackson}, y haciendo uso de esta se puede reescribir (\ref{SistemaLinealNohomogeneo}) como:
\begin{equation}\label{PotentialOndas}
\begin{cases} 
\vspace{0.2cm}\displaystyle\frac{1}{c^2}\frac{\partial^2 {\bf A_L}}{\partial t^2}- \Delta {\bf A_L}=\mu{\bf j},\,\,\,\,\,\,\,\,\,\cr\displaystyle\frac{1}{c^2}\frac{\partial^2 V_L}{\partial t^2}- \Delta V_L=\frac{\rho}{\varepsilon}.
\end{cases}
\end{equation}
Note que (\ref{PotentialrelationPrime1}) y (\ref{PotentialOndas}) no determinan un único par $({\bf A_L},V_L)$ cuando $\bf j$ y $\rho$ son conocidas.

\begin{figure}[H]
\centering
\includegraphics[width=8.5cm,height=6.5cm]{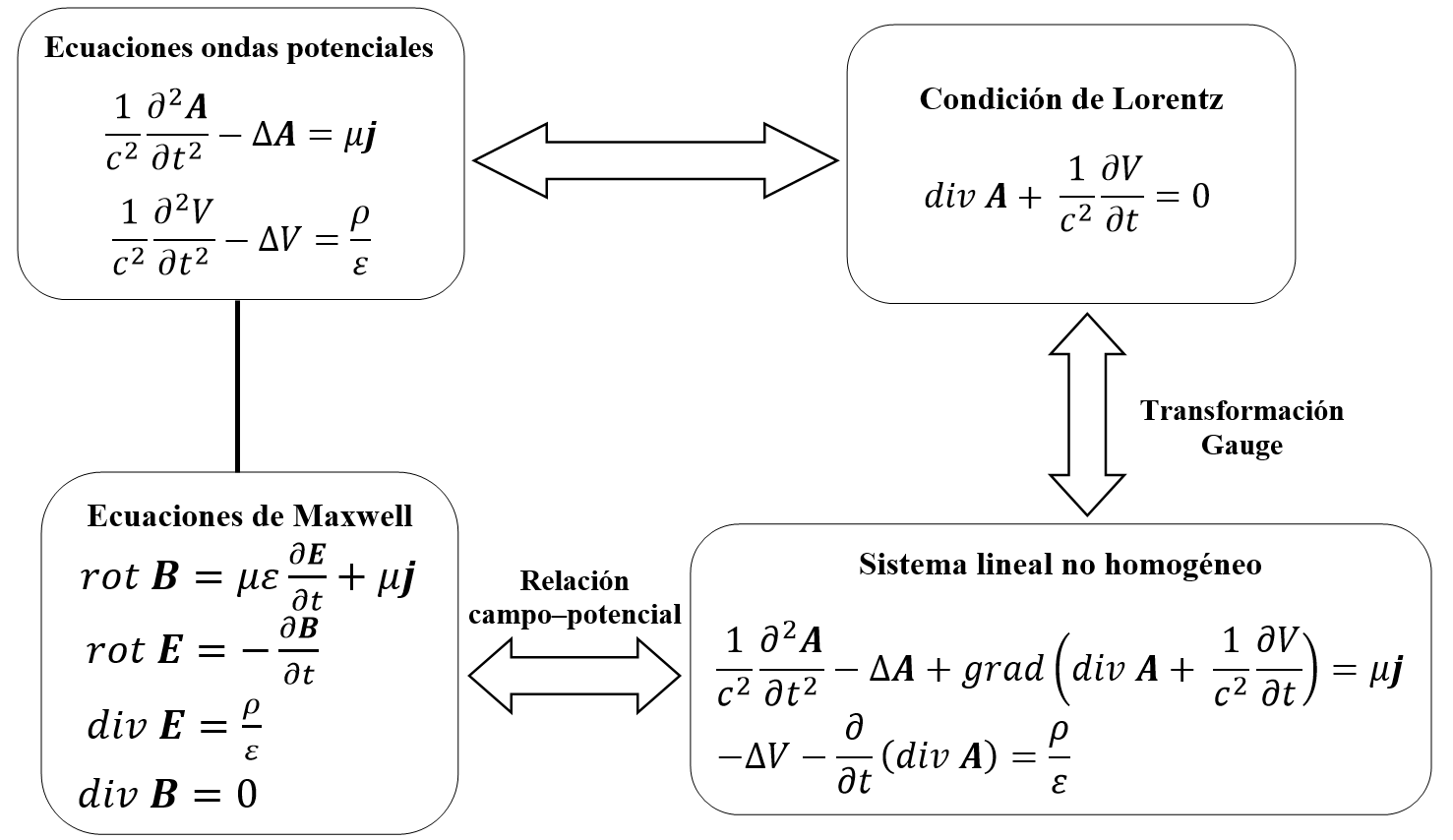}
\caption{Condición de Lorentz y niveles de descripción.}
Fuente: Elaboración propia 
\label{ModeloPropEmergentes2}
\end{figure}

En la Figura \ref{ModeloPropEmergentes2} se representa la condición de Lorentz como propiedad emergente de las ecuaciones de Maxwell. Se considera el paso entre las ecuaciones de Maxwell y la condición de Lorentz, lo que puede ser representado en dos niveles. Inicialmente, en el nivel inferior izquierdo se encuentran las ecuaciones de Maxwell, luego utilizando las relaciones de los campos electromagnéticos con los potenciales escalar y vectorial, obtenemos el sistema lineal no homogéneo que aparece en el nivel inferior derecho. Para el paso al nivel superior es necesario agregar información adicional, y así del sistema lineal no homogéneo se deduce la condición de Lorentz, la cual permite representar el sistema a través de ecuaciones de ondas que involucran los potenciales (nivel superior izquierdo) \cite{Fuentes}. 

Por lo anterior, la condición de Lorentz en una propiedad emergente del sistema de ecuaciones de Maxwell, ya que surge al considerar los elementos (ecuaciones) y sus interrelaciones (sustituciones convenientes de unas ecuaciones en otras). Resultaría imposible obtener la condición de Lorentz al considerar cada ecuación del sistema de Maxwell de manera individual. 
  
En el caso de soluciones monocromáticas (armónicas en el tiempo) del sistema de Maxwell, consideremos las expresiones tiempo-armónicas de los campos
\begin{equation}\label{PotentialTimeHarmonic}
\begin{cases} 
\vspace{0.1cm}{\bf B}(\vec{x},t)=\vec{B}(\vec{x})e^{i\omega t},\,\,\,\,\,\,\,\,\,\cr{\bf E}(\vec{x},t)=\vec{E}(\vec{x})e^{i\omega t}.
\end{cases}
\end{equation}
De manera análoga, consideremos las expresiones tiempo-armónicas de los potenciales escalar y vectorial, y usando dichas expresiones obtenemos de (\ref{Potentialrelation}) las ecuaciones
\begin{equation}\label{PotentialRelationTimeHarmonic}
\begin{cases} 
\vspace{0.1cm}\vec{B}(\vec{x})=\textup{rot}\,\vec{A}(\vec{x}),\,\,\,\,\,\,\,\,\,\cr\vec{E}(\vec{x})=-\textup{grad}\,V(\vec{x})-i\omega \vec{A}(\vec{x}).
\end{cases}
\end{equation}
Escribamos estas relaciones en el lenguaje de los cuaternios. Así, interpretamos los potenciales escalar y vectorial como un paravector especial $\vec{F}(\vec{x})=iV(\vec{x})+\vec{A}(\vec{x})$ y el campo eléctrico y la inducción magnética como la función cuaterniónica $\vec{U}(\vec{x})=-i\vec{E}(\vec{x})+\vec{B}(\vec{x})$ y escribimos 
\begin{equation}\label{FuncionCuaternionicaU}
\vec{U}(\vec{x})=\left(D-i\omega\right)\vec{F}(\vec{x}).
\end{equation}
Considerando además las expresiones tiempo-armónicas de la densidad de corriente y densidad de carga, las ecuaciones (\ref{PotentialOndas}) para el potencial escalar y vectorial son transformadas en
\begin{equation}\label{PotentialHelmholtz}
\left(\Delta+\kappa^2\right)\vec{F}(\vec{x})=\vec{R}(\vec{x}),
\end{equation}
para $\kappa=\displaystyle\frac{\omega}{c}$ y $\vec{R}(\vec{x})=\left(-i\displaystyle\frac{\rho(\vec{x})}{\varepsilon},-\mu\vec{j}(\vec{x})\right)$.\\
En términos del operador de Dirac, (\ref{PotentialHelmholtz}) es equivalente a
\begin{equation}\label{PotentialFactorization}
\left(D+\kappa\right)\left(D-\kappa\right)\vec{F}(\vec{x})=-\vec{R}(\vec{x}).
\end{equation}
Así, la condición de Lorentz permite obtener una representación del sistema de Maxwell tiempo armónico en el contexto cuaterniónico, pero utilizando los potenciales escalar y vectorial \cite{Bernstein}.

\begin{figure}[H]
\centering
\includegraphics[width=8cm,height=8cm]{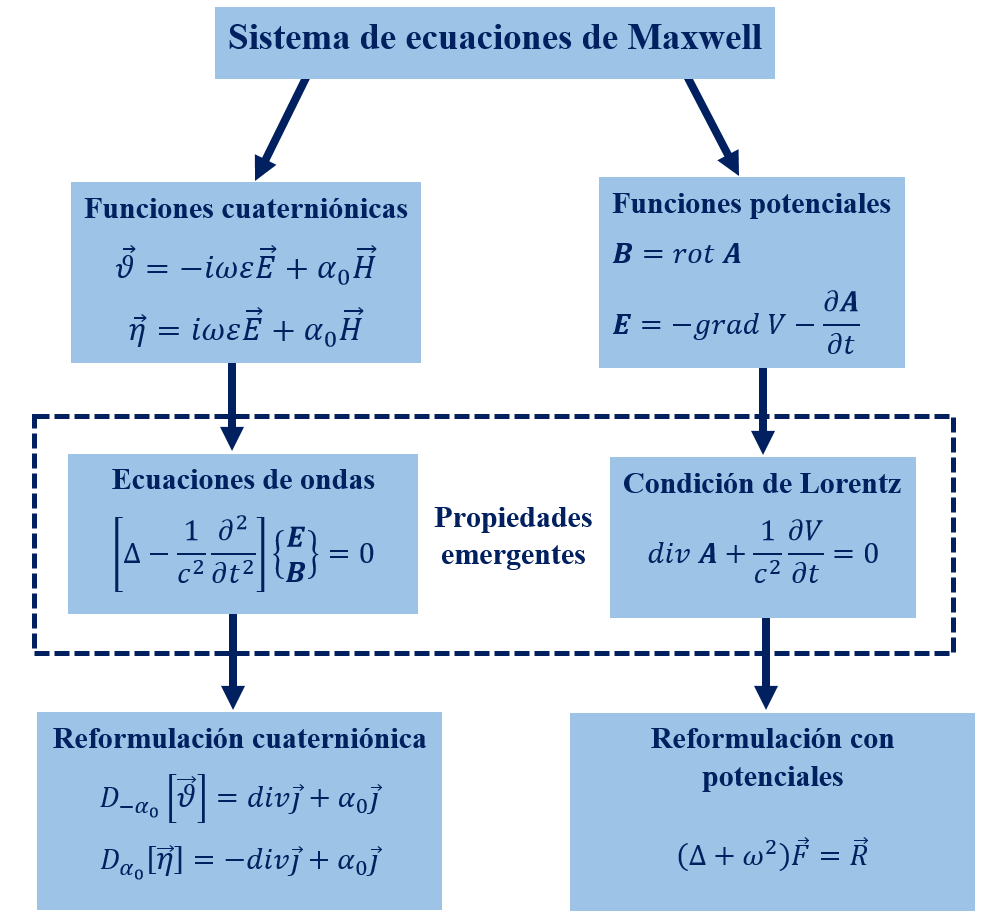}
\caption{Modelo lineal: reformulaciones del sistema Maxwell.}
Fuente: Elaboración propia 
\label{Reformulaciones}
\end{figure}

Los dos métodos anteriores son representados en la Figura \ref{Reformulaciones} por un modelo lineal (organigrama) con dos ramas, cada una de las cuales finaliza en una reformulación del sistema de ecuaciones de Maxwell tiempo armónico, utilizando en cada caso una de las dos propiedades emergentes estudiadas. 
La rama de la derecha muestra el camino por el que se obtiene una reescritura de las ecuaciones de Maxwell a través de una ecuación de Helmholtz y en términos de potenciales \cite{Bernstein}. En un primer nivel aparecen dos ecuaciones que enuncian los campos electromagnéticos en términos de los potenciales escalar y vectorial. Sin embargo, sucede que los potenciales escalar y vectorial no determinan unívocamente a los elementos que tienen un significado físico directo (el campo eléctrico y magnético), y por tanto, no se les asigna un significado físico directo a los potenciales. Así, en el nivel intermedio de la rama, la condición de Lorentz fija esa arbitrariedad para elegir los potenciales escalar y vectorial, lo que finalmente permite obtener dicha reformulación. 
Por su parte, la rama de la izquierda ilustra el camino para obtener una reformulación utilizando el operador de Moisil-Teodoresco aplicado sobre funciones cuaterniónicas complejas \cite{Kravchenko}. En este caso, en el primer nivel se definen las dos funciones cuaterniónicas complejas en términos de los campos electromagnéticos. En el nivel intermedio, intervienen las ecuaciones de ondas, que en el caso tiempo-armónico se convierten en ecuaciones tipo Helmholtz. Estas motivaron la introducción del operador de Helmholtz, el cual puede ser factorizado por el operador de Dirac desplazado, que resulta clave para finalmente obtener la reformulación de las ecuaciones de Maxwell.

\section{Conclusiones}
En este artículo, con base a los conceptos de emergencia según la teoría general de sistemas se consideraron propiedades emergentes del electromagnetismo, en particular, la condición de Lorentz y las ecuaciones de ondas electromagnéticas. Además, haciendo uso del Análisis Cuaterniónico y de las relaciones entre los campos electromagnéticos y los potenciales escalar y vectorial, se muestra la aplicación e importancia de estas propiedades que emergen en el Sistema de Ecuaciones de Maxwell en un contexto sistémico.

\section*{Agradecimientos} 
Yudier Peña Pérez agradece el apoyo financiero del Consejo Nacional de Ciencia y Tecnología (CONACYT) mediante una beca de estudios de posgrado (número de becario 744134). Juan Bory Reyes fue parcialmente apoyado por el Instituto Politécnico Nacional en el marco de los programas SIP (SIP20200363).


\end{document}